# TYPE III DYSON SPHERE OF HIGHLY ADVANCED CIVILISATIONS AROUND A SUPER MASSIVE BLACK HOLE

**MAKOTO INOUE[1] AND HIROMITSU YOKOO[2]**
1. *Institute of Astronomy and Astrophysics, Academia Sinica, P.O Box 23-141, Taipei 10617, Taiwan, R.O.C.*
2. *Chiba University of Commerce, 1-3-1 Konodai, Ichikawa-shi, Chiba, 272-8512, Japan.*
Email: inoue@asiaa.sinica.edu.tw[1]

We describe a new system for a society of highly advanced civilizations around a super massive black hole (SMBH), as an advanced Type III "Dyson Sphere," pointing out an efficient usage of energy for the advanced civilizations. SMBH also works as a sink for waste materials. Here we assume that Type III civilisations of Kardashev classification [1] form a galactic club [2] in a galaxy, and the energy from the SMBH will be delivered to the club members, forming an energy control system similar to power grids in our present society. The energy is probably transmitted by a sharp beam with coherent electro-magnetic waves, which provide a new concept for the search for extraterrestrial intelligence (SETI) via detection of such energy transmission signals. This expands the search window for other intelligences within the Universe.

**Keywords:** Type III Dyson Sphere, Super Massive Black Hole, Energy transmission, Galactic club, SETI

## 1. INTRODUCTION

Through the evolutionary process of the Universe, it is a great mystery to create life and develop it to highly advanced intelligence. Motivated by this mystery, the search for extraterrestrial intelligence (SETI) has been conducted for a long time, mainly by radio signals with negative results so far. Drake recently reviewed this issue [3]. The SETI project has aimed at stars around which planets are expected to be a cradle of life for developing intelligence and advanced technological civilizations.

Dyson Sphere is anticipated as a growing stage of advanced civilisations [4]. In this concept, a star is the expected energy source to support the activity for a civilisation which has developed around it. Such a system provides a possible means for energy acquisition by a highly civilised intelligence who should require a prodigious amount of energy to develop and maintain the civilisation. After using energy, radiation energy from the central star is converted and radiated to lower frequency radiation, e.g., as infrared radiation, from the Dyson Sphere. Jugaku and Nishimura conducted intensively the searches for Dyson Spheres ([5], references therein). Bradbury revisited this concept [6].

Kardashev mentioned Type III civilizations as the most advanced stage, which controls the energy on a galactic scale [1]. He classified civilizations in three types depending on the technology level with energy consumption which increases with the development of technology level:

> Type I – technological level close to that of the present one, with energy consumption at ~ $4 \times 10^{19}$ erg/s.
>
> Type II – a civilization capable of harnessing the energy radiated by its own star (i.e., the stage of "Dyson Sphere"), with energy consumption at ~ $4 \times 10^{33}$ erg/s.
>
> Type III – a civilization in possession of energy on the scale of its own galaxy, with energy consumption at ~ $4 \times 10^{44}$ erg/s.

However, the energy management was not well discussed.

The purpose of this paper is to study a new system for a society of highly advanced intelligences around a super massive black hole (SMBH), as an advanced Type III, pointing out an efficient usage of energy for the advanced civilizations. The system could provide the intelligences a distinct advantage for supporting the higher activities in their society and further possible developments. In Section 2, a possible advanced system is described. In Section 3, a new SETI capability inherent to this system is discussed.

## 2. ADVANCED SYSTEM

A society of a highly advanced civilization is supposed to require a huge energy to operate the social system. As the gravitational energy released by the accretion of matter onto a SMBH is huge, a system must be developed to use this energy in such a society. The condition around a SMBH at the centre of galaxy would be more efficient both in extracting energy and exhausting the waste energy for advanced civilizations, than those of a Dyson Sphere [4]. Some active galactic nuclei (AGN) are extremely luminous, hundreds of times the integrated stellar luminosity of a whole galaxy. For example, bolometric luminosity of QSOs and Seyfert galaxies is distributed mostly in a range between $10^{43}$-$10^{47}$ erg/s [7]. Mean spectral energy distributions of QSOs and Blazars are shown in Fig. 3 of Sanders & Mirabel [8]. The energy is derived from the gravitational energy of the central SMBH. A huge amount of radiation is generated in a close vicinity of SMBH, where an accretion disk is rotating around SMBH, and the potential energy of the accreting matter is released to form a hot and dense disk. The size of SMBH is typically measured in units of the Schwarzschild radius $Rs = 2GM/c^2$, where G and c are the gravitational constant and the velocity of light, respectively, and $M$ is the mass of the SMBH. In a case of $M = 10^8$ $M_\odot$, $Rs$ is only 2 AU (1 AU = $1.50 \times 10^8$ km; is the radius of the Earth orbit). The diameter of an accretion disk is likely the size of our solar system, comparable to Dyson Sphere of stellar scale Type II civilizations [1], while the energy scale is more than $10^8$ times larger than that, i.e., the integrated stellar energy of the whole galaxy.

Radiation from the accretion disk will be collected by a mirror system as a Type III "Dyson Sphere." Waste material and energy could be thrown off toward the central SMBH, and the SMBH would be the final reservoir for all of the waste materials for any civilizations. Thus, the most advanced civilizations would develop their activities using a SMBH efficiently, putting the power plants around the SMBH at the centre of their home galaxy. We discuss a possible model of such system, and the possibility to detect indicators of the existence of the system, or such civilisations.

### 2.1 Power Plant around SMBH

The structures of the power plant basically revolve around the central SMBH in Keplerian motion to form "Dyson Shells." In an advanced case of Type II, the central star is almost fully covered to form "Dyson Sphere" (see [6]). Here we discuss the case of structures partly covered, or the Dyson Shell type, and call it a Dyson Sphere. Unlike a stellar environment, or Type II Dyson Sphere, there are complex structures like relativistic jets, accretion disk and accreting matters, rapidly rotating stars, etc., and hence it would be very difficult to construct a fully covered structure, like a system studied by Birch over a large gaseous planet (e.g., Jupiter) [9]. However, it is not easy to set numbers of power plants with similar distance orbiting around the central SMBH. Hence, it would be a possible solution to set the power plants on a solid framework, something like structures studied by Birch [9]. Some areas should be kept uncovered to yield emanating jets and accreting flows.

First, we estimate the distance where iron melts by radiation, assuming $10^{45}$ erg/s comes from a central point source. Given this luminosity is generated by 10% of the Eddington luminosity, the SMBH mass corresponds to $7 \times 10^7$ M$_\odot$. The melting point of iron is $1.8 \times 10^3$ K, and at 0.75 pc ($2.3 \times 10^{13}$ km) from the central source, temperature becomes the melting point. It would be possible to use material with a high reflection coefficient (i.e., large albedo) and high melting point, and as the distance is inversely proportional to temperature square, the area of 0.1 pc ($3 \times 10^{12}$ km) from SMBH would be available to construct the structures. This area corresponds $\sim 10^4$ $R$s for $M \sim 10^8$ M$_\odot$.

The power plants will be set off-plane of the accretion disk or outside of the outer edge of it to avoid its orbit crossing the disk. Then, the distance from the central SMBH might be 1,000 $R$s or more, and the tidal force should not be a serious problem for the structure of the power plant. As the tidal force at a distance $R$ from a SMBH (gravity centre) is inversely proportional to $R^3$, and $R$s is proportional to $M$, the tidal force decreases with mass of SMBH. For example, the tidal force at 10 $R$s is already $10^{-3}$ times smaller than that at the surface of the Earth, for the case of $10^8$ M$_\odot$. As the power plants are supposed to set far out from 10 $R$s and the tidal force decreases with $R^{-3}$ from the central SMBH, we will not care about the tidal disruption of the structure in the power plants anymore.

On the other hand, in the areas outside of the accretion disk, the power plants may suffer from interaction with stars orbiting around the SMBH. They might be scattered from the original orbit unless they are actively controlled. The closest orbit of a star rotating around Sgr A*, the SMBH in our Galactic Centre, was at 600 $R$s [10]. A possible orbit would be compromised by the outer extent of the accretion disk and innermost orbit of stars. The proper area for the power plants would be around $10^3$-$10^4$ $R$s, although parameters of the outer edge of accretion disk and innermost stellar orbit are not well understood.

It is interesting to investigate the energy efficiency at the power plant for further consideration. The dominant energy flow is the outgoing radiation from the accretion disk. Some fraction of the radiation energy is reflected and focused onto the power plant. The energy efficiency could be a measure for a level of civilisation.

### 2.2 Energy Transmission

In a galactic club (e.g., [11]), habitants are not necessarily living at the same place as the power plant near the SMBH. The power plant could be even remotely controlled from the intelligent habitat. In this case, the energy transmission from the power plant to the habitat should be accomplished presumably by electro-magnetic radiation. Selection of the wavelength region is a key in making an efficient energy transmission. The photon energy is higher for shorter wavelengths, while absorption and scattering are stronger at shorter wavelengths. Recently, many gamma-ray sources have been detected in AGNs (e.g., [12]). However, as the matter density in the central region of galaxy is higher, absorption and scattering are greater and the energy transmission at gamma-ray wavelengths is not so efficient. In terms of transparency, the near infrared region would be a possible wavelength. Stars rotating around the Galactic Centre are observed ([13], references therein), although Sgr A* is still difficult to observe at infrared wavelengths. This suggests that when we choose the transmission frequency, we must take account of (1) the spectral energy distribution of the radiation from the accretion disk, (2) energy conversion efficiency of the power plant for the transmission, and (3) transmission efficiency through the interstellar medium of the galaxy.

In addition to the selection of the wavelength region, techniques of focusing the transmitting beam should take account of the efficiency of energy transmission. In the present techniques, we can reduce the observing beamwidth down to several tens of micro arc-seconds using Very Long Baseline Interferometry (VLBI) at the sub millimetre wavelength region, which is revealing the fine structure of Sgr A* [14]. The VLBI technique combines to produce a coherent reception by several separate antennae. Because of the reciprocity between reception and transmission, we can send a very sharp beam by coherent radiation with separate transmitters or antennae. The beamwidth is $\approx \lambda/D$, where $\lambda$ and D are the wavelength and the separation between transmitting antennae, respectively. Hence, the beamwidth is directly related to the selection of $\lambda$. The efficient wavelength should first be selected and the separation D of the transmitter is then optimized for the target areas to transmit the energy. In the coherent transmission, one must take into account the phase disturbance by the interstellar medium, in addition to its absorption, for the frequency selection.

Instead of VLBI array, one filled aperture of the transmitter is efficient for transmitting a large energy, in terms of number of transmitters for each array element and the coherent system. The construction of a large filled aperture may not be easy. However, the advanced technology would make it possible to send the required energy by a single large aperture. At $\lambda \approx 1$ µm, the 1 µ arc-second beam requires an aperture of 200 km in diameter. In this system, the beam has the same size of the aperture in the near field region, and the beam gets into the far field region at around 1 pc, and the transmitted power flux density is reduced by 80 dB at 10 kpc ($3 \times 10^{17}$ km). This reduction is still acceptable for delivering the energy within a galaxy from a SMBH, instead of a local star, assuming the efficiency available from Dyson Sphere, or a star, and that from SMBH is the same. The difference of the generated energy between star and SMBH is more than 80 dB.

When a molecule in a gas cloud is pumped to a population inversion state, this gas cloud can be used as an efficient transmission system, by means of maser or laser emission mechanism. The seed radiation is amplified coherently and the resultant radiation has a sharp beam. For example, in a case of water maser emission at $\lambda \approx 1.3$ cm, a gas cloud of $3 \times 10^6$ km (0.02 AU) has the same beam width of 1 µ arc-second mentioned above. For the collisional pumping the $H_2$ gas density of $10^8 \sim 10^9$ cm$^{-3}$ is required [15]. A part of molecular torus rotating around the accretion disk could be used to make the population inversion state artificially. In fact, mega-maser emissions rotating around galactic centers have been observed [16], and the artificial pumping may not be so difficult. Some mega-maser emissions have isotropic luminosities > 100 $L_\odot$ [17]. When the beamwidth could be shaped into a 10-arcsec level, the energy of $10^{45}$ erg/s would be transferred, given the isotropic radiation. However, as the bandwidth of the maser emission is narrow, a frequency-spread system with wide bandwidth might be applied to transmit a huge amount of energy more efficiently.

Even using a huge aperture at short wavelength, it may not be enough to supply the total required energy for a civilized society living a distant area from the power plant. The commonly used energy source is probably nuclear power from local power stations, when necessary. The nuclear reactors could be either artificial or natural (i.e., stars), within the territory of each civilized society. These power stations would supply the optional power to each user society.

Highly beamed radiation with high flux density is useful for sailing in interstellar space, and such spacecrafts may be supplied by the energy transmitted from the central power plant, or a local power station nearby. The spacecraft could be installed with one of several types of energy suppliers to sail interstellar space. One energy supplier will be a nuclear power. The other energy supplier would be the transmitted energy from the power plant, or local power stations described above.

The energy transmission system in a galactic club probably forms power grids within a galaxy. At the central power stations, the direction of power transmission will be switched from time to time due to different galactic rotation between transmitter and recipient. Then, a properly coordinated system like a power grid system is required. Further, when a power plant fails, it should be reconfigured soon as it takes long time to get the information from a recipient who might be located some kpc (1 kpc = 3260 light year = $3.08 \times 10^{16}$ km) from the center of galaxy. Such local civilisations probably have also a back-up system to form a power grid as well among recipients. As it takes time to communicate with each other, an efficient communication system in the galactic club would be formed in an inner area of galaxy.

## 3. DETECTION OF POWER PLANTS

It is difficult to detect the signal of the energy transmission, as it is highly beamed and the chance for detection is low. Chance coincidence of detection for a 1 µ arc-second beam is less than $10^{-23}$, simply to find the beam by chance. This probability for the isotropic case increases when a galaxy is seen edge on, assuming the galactic club members lie within the galactic plane of spiral galaxies. Also, the detection probability increases when the energy transmission is made by many power plants using a large aperture, or a multi beam system from the power plants. However, in any case, the beam comes from the direction of the central region around the SMBH, and careful discrimination is required between direct radiation from the SMBH region and that from the power plants. One key point is the coherence of the signal, just like maser emissions. Maser observations were proposed as a probe of interplanetary communications from advanced intelligences [18]. Maser surveys for galaxies have been made (e.g., [16]), and the Keplerian rotation of a system of maser components was in fact discovered [19]. In this case, the innermost radius of the maser components is $3 \times 10^4$ times of its $R$s. The maser radiation, however, does not seem unnatural, although the system is not fully understood: For example, why are the main maser components in front of the SMBH all aligned almost on the same orbit?

The energy flux density of the energy transmission should be much higher than that of the communication signal which SETI programs have been assuming so far, and the frequency is not necessarily the same as the natural maser emissions. Advanced technologies will make it possible to create a high power coherent oscillator with broad bandwidth. Unnatural maser-like emission, or coherent emission, from the central SMBH region of galaxy could correspond to a signature of energy transmission from the power plant by an advanced civilization. In addition, as Kardashev mentioned [20], large artificial structures could be searched against SMBH. The mirror system designed to reflect and transmit the energy could be detected as a shadow against the bright accretion disk around the SMBH. Also, the maser-like emission from local power stations could be observed if a galactic club has developed a power grid system in a galaxy as envisioned here.

## 4. CONCLUDING REMARKS

Figure 1 shows a schematic picture of the power plants and others around SMBH. The condition around SMBH is very promising for an advanced intelligence to manage energy issues in terms of both energy generation and disposition. This idea comes from a combination of Type III and Dyson Sphere. The strong radiation from the accretion disk rotating around SMBH is mainly used, and the waste energy is returned toward the SMBH. The available energy is huge compared to Dyson Sphere of stellar scale, and this type of civilization could be called Type III Dyson Sphere. The search for this type of civilization, however, would not likely be revealed from "unintended" communication signal, but detection by coherent radiation from power stations may be more promising.

## ACKNOWLEDGEMENTS

We thank R. Taam for comments to the manuscript and Y.C. Tsai for the picture.

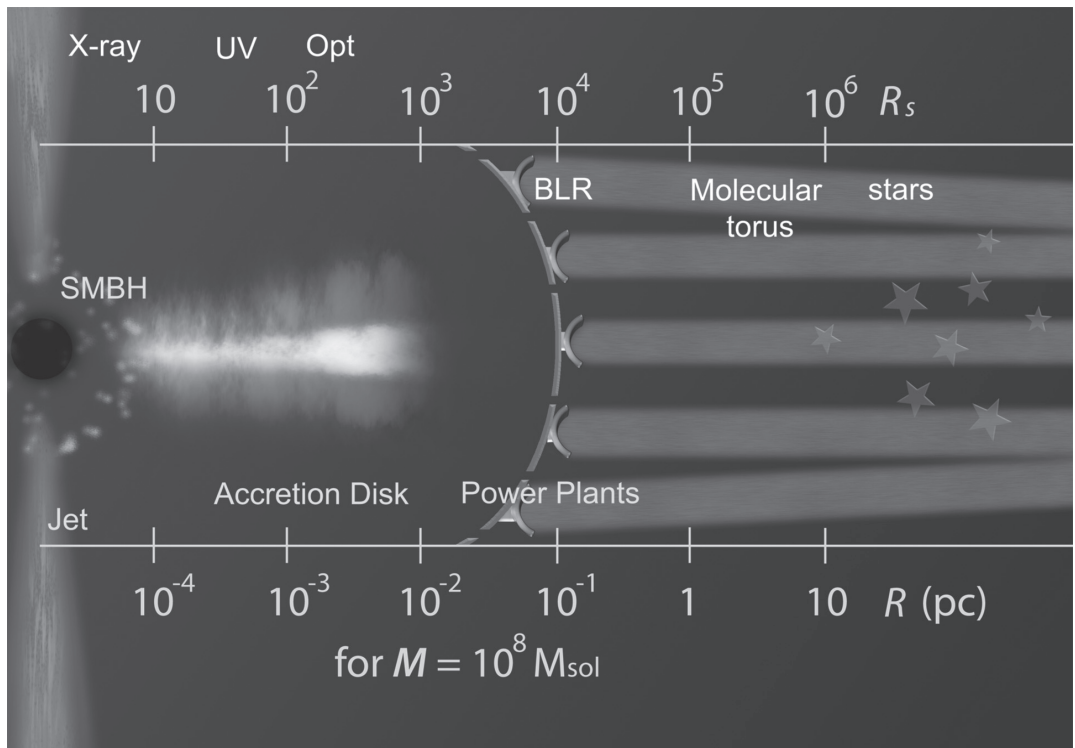

**Fig. 1** Schematic picture around SMBH. Items are not to scale. In this picture, an example of power plants with transmitters is shown partly. BLR stands for the Broad Line Region. The SMBH and accretion disk will not be fully covered by the collectors of power plants, so as not to prevent jets emanating from somewhere in this area, and accretion flow coming out of the central region. A twin jet is thought to emanate perpendicular to the plane of accretion disk, seen about 10% of AGN. The energy from the power plants is transferred by electro-magnetic waves to habitats of advanced civilizations. In this picture, the beams are directed to a galactic plane on which a galactic club is formed. However, the planes of the accretion disk and the host galaxy may not necessarily be in the same plane (e.g., Inoue [21]). One $R_s$ for the SMBH mass of $10^8$ $M_\odot$ is $\sim 10^{-5}$ pc ($3 \times 10^8$ km).

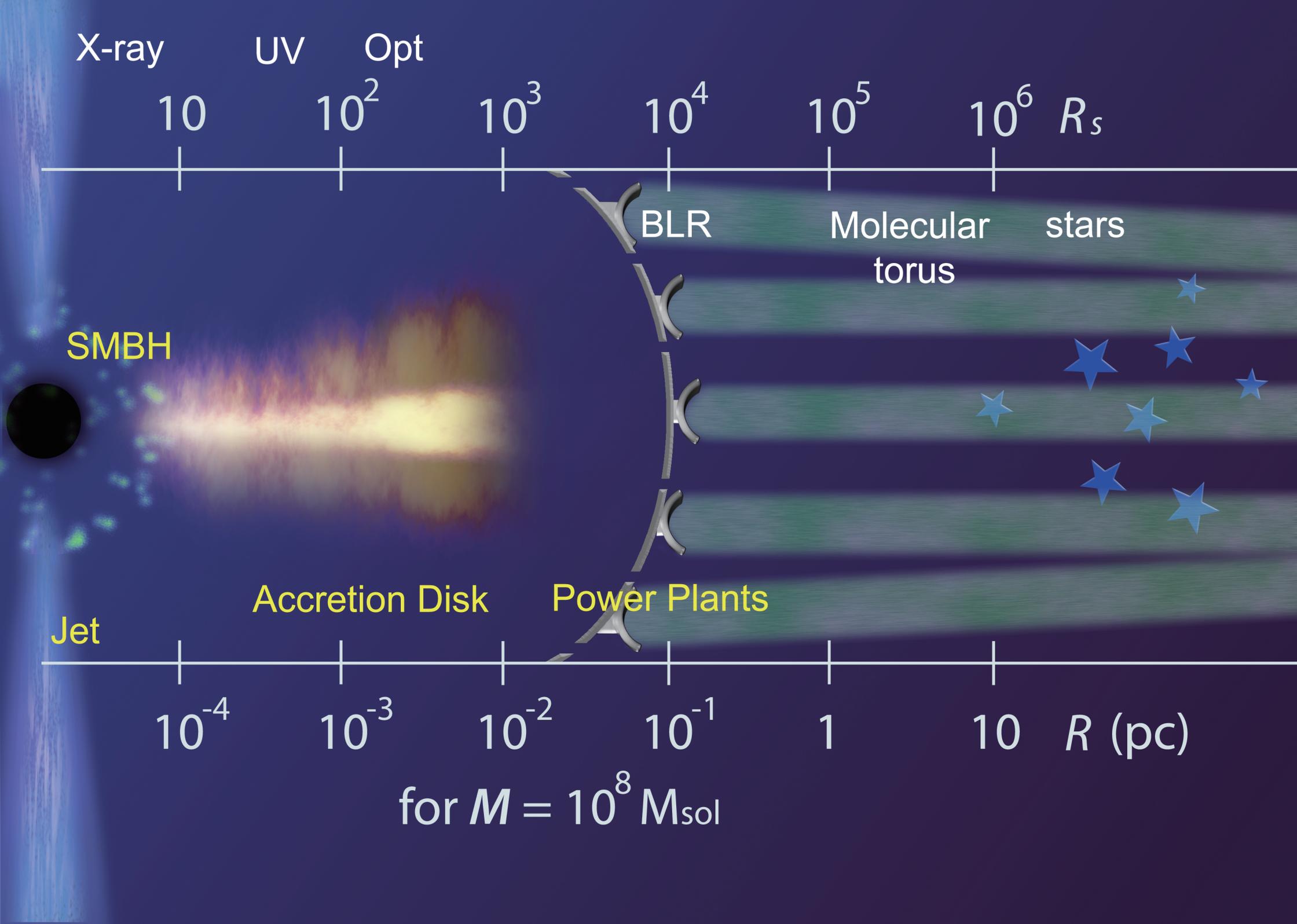